\documentstyle[12pt,epsfig]{article}
\input epsf

\begin{document}

\begin{flushright}
RUB-TP2-20/00
\end{flushright}
\begin{center}

{\bf {\Large
Meson twist-4 parton distributions in terms of twist-2 distribution
amplitudes at large $N_{c}$}}\\[0.3cm]

N.-Y.~Lee$^a$,
P.V.~Pobylitsa$^{a,b}$, M.V.~Polyakov$^{a,b}$, and K. Goeke$^a$
\\[0.3cm]
\normalsize $^a$Institut f\"ur Theoretische Physik II,
Ruhr--Universit\"at Bochum,\\ \normalsize D--44780 Bochum, Germany\\
\normalsize $^b$ Petersburg Nuclear Physics Institute, Gatchina, \\
\normalsize St.Petersburg 188350, Russia
\end{center}


%
\begin{abstract}
\noindent
We show that
in the large $N_c$ limit four-quark twist-4 distributions in the
pion can be expressed in terms of twist-2 pion distribution
amplitude. This allows us to compute the isospin-2 structure function
of the pion $F_2^{I=2}(x_B)$ in the large $N_c$ limit.
The method can be easily applied to other mesons as well.
\end{abstract}

\noindent
{\bf 1.}~Matrix elements of four-quark operators
appear {\it e.g.}\ in twist-4 contributions
to the structure functions of deep--inelastic
scattering \cite{Shuryak:1982kj,Jaffe:1981td}, or
in the description of non-leptonic weak
decays \cite{Shifman:1977ge}.
In this paper we consider the specific example of
isospin-2 pion structure functions $F_2(x_B)$
\begin{equation}
F_{2}^{I=2}=F_{2}^{\pi
^{+}}+F_{2}^{\pi ^{-}}-2 F_{2}^{\pi ^{0}}\, .
\label{pervoe}
\end{equation}
In this particular isospin combination of structure functions the
contribution of twist-2 operators cancels out exactly and
twist-4 contributions are dominant.
It is remarkable that in the large $N_{c}$ limit the twist-4
distribution function of quarks in the pion can be reduced to a convolution
of twist-2 pion distribution amplitudes.

        \noindent
{\bf 2.}~One can read
the contribution of four-quark twist-4 operators to $F_{2}$
from \cite{EFP,SQ} (correcting obvious misprints)

\[
\frac{1}{x_{B}}F_{2}^{twist-4}(x_{B})=\frac{1}{Q^{2}} \int
dx_{1}dxdx_{2}\,x_{B}
\]
\[
\times \left\{ \left[ \Delta (x_{2},x_{1},x)+\Delta (y_{2},y_{1},x)\right]
T_{qq}^{(+)}(x_{1},x,x_{2})\right.
\]
\begin{equation}
\left. -\left[ \Delta (x_{2},y_{1},x)+\Delta (y_{2},x_{1},x)\right]
T_{qq}^{(-)}(x_{1},x,x_{2})\right\}\, .
\label{F2-tw-4-corrected-start-factorized}
\end{equation}
Here $y_1\equiv x-x_1,\ y_2\equiv x-x_2$.
We use the function $\Delta$ from ref.~\cite{EFP}

\begin{equation}
\Delta (x_{1},x_{2},x_{3})=-\frac{1}{\pi }{\rm Im}
\Biggl[\frac{1}{x_{1}-x_{B}+i0}\,
\frac{1}{x_{2}-x_{B}+i0}\,\frac{1}{x_{3}-x_{B}+i0}
\Biggr] \, .
\label{Delta-def-1}
\end{equation}
The functions $T_{qq}^{(\pm )}(x_{1},x,x_{2})$ are defined as
follows
\begin{equation}
T_{qq}^{(\pm )}(x_{1},x,x_{2})=T_{qq}(x_{1},x,x_{2})\pm \tilde{T}
_{qq}(x_{1},x,x_{2})  \label{T-plus-def}
\end{equation}
with $T_{qq}$, $\tilde{T}_{qq}$ given by

\[
\left.
\begin{array}{c}
T_{qq}(x_{1},x,x_{2}) \\
\tilde{T}_{qq}(x_{1},x,x_{2})
\end{array}
\right\} =g^{2}\frac{p^+}{4}
\int \frac{dz_{1}^{-}}{2\pi }\,\frac{dz_{2}^{-}}{
2\pi }\,\frac{dz^{-}}{2\pi }
\]
\[
\times \exp \left\{ ip^{+}\left[
x_{1}z_{1}^{-}+(x-x_{1})z^{-}-(x-x_{2})z_{2}^{-}\right] \right\}
\]
\begin{equation}
\times \langle p|T\left\{ \left[ \bar{\psi}(0)\Gamma t^{B}\psi (z_{2}^{-})
\right] \left[ \bar{\psi}(z^{-})\Gamma t^{B}\psi (z_{1}^{-})\right] \right\}
|p\rangle \, . \label{T-T-tilde-def}
\end{equation}
Here $g$ is the QCD coupling constant,
$t^B=\frac 12 \lambda^B$ are generators of the color group and
\begin{equation}
\Gamma =\left\{
\begin{array}{cc}
\gamma^+Q_{e} & {\rm for\ }T_{qq} \\
\gamma _{5}\gamma^+Q_{e} & {\rm for\ }\tilde{T}_{qq}
\end{array}
\right. \, ,
\end{equation}
where $Q_e$ is the quark electric charge matrix. The light-cone
components of a vector $V^\mu$ are defined as $V^\pm=\frac{1}{\sqrt{2}}(V^0\pm
V^3)$.

\noindent
{\bf 3.}~Generically the meson matrix elements of four-quark twist-4 operators
are not related to matrix elements of two-quark operators. Here we
show that in the limit of large number of colors $N_c$ the matrix
elements of four-quark operators can be expressed in terms of marix
elements of two-quark operators. This observation allows us to
express twist-4 contribution to $F_2(x)$ of the pion in terms of
twist-2 distribution amplitudes.

A simple analysis of leading in $N_c$  Feynman diagrams
\cite{tHooft-74},
corresponding to the meson matrix element
\begin{equation}
\langle M_{1}|(\bar{\psi}O_{1}\psi )(\bar{\psi}O_{2}\psi )|M_{2}\rangle
\label{me}
\end{equation}
with matrices $O_i$ being unity in the color subspace,
shows that this matrix elements can be factorized in the limit of
large number of colors as

\[
\langle M_{1}|(\bar{\psi}O_{1}\psi )(\bar{\psi}O_{2}\psi )|M_{2}\rangle
\stackrel{N_{c}\rightarrow \infty }{\longrightarrow}
\langle
M_{1}|M_{2}\rangle \langle 0|(\bar{\psi}O_{1}\psi )(\bar{\psi}O_{2}\psi
)|0\rangle
\]
\begin{equation}
+\langle M_{1}|(\bar{\psi}O_{1}\psi )|0\rangle \,\langle 0|(\bar{\psi}
O_{2}\psi )|M_{2}\rangle +\langle M_{1}|(\bar{\psi}O_{2}\psi )|0\rangle
\,\langle 0|(\bar{\psi}O_{1}\psi )|M_{2}\rangle \, .
\label{Factorization-M-1}
\end{equation}
Let us define the connected
part of the matrix element as follows
\[
\langle M_{1}|(\bar{\psi}O_{1}\psi )(\bar{\psi}O_{2}\psi )|M_{2}\rangle
_{connected}
\]
\begin{equation}
\equiv \langle M_{1}|(\bar{\psi}O_{1}\psi )(\bar{\psi}O_{2}\psi
)|M_{2}\rangle -\langle M_{1}|M_{2}\rangle \langle 0|(\bar{\psi}O_{1}\psi )(
\bar{\psi}O_{2}\psi )|0\rangle \, .
\end{equation}
If mesons
$M_{1}$ and $M_{2}$ have different momenta or different internal quantum
numbers then $\langle M_{1}|M_{2}\rangle =0$ and there is no difference
between the matrix elements and its connected part. But if $|M_{1}\rangle
=|M_{2}\rangle $ then the nonconnected matrix element $\langle M_{1}|(\bar{
\psi}O_{1}\psi )(\bar{\psi}O_{2}\psi )|M_{2}\rangle $ is divergent and we
must consider the connected part.
Now the factorization (\ref{Factorization-M-1}) takes the form
\[
\langle M_{1}|(\bar{\psi}O_{1}\psi )(\bar{\psi}O_{2}\psi )|M_{2}\rangle
_{connected}
\]
\begin{equation}
\stackrel{N_{c}\rightarrow \infty }{\longrightarrow }\langle M_{1}|(\bar{\psi}
O_{1}\psi )|0\rangle \,\langle 0|(\bar{\psi}O_{2}\psi )|M_{2}\rangle
+\langle M_{1}|(\bar{\psi}O_{2}\psi )|0\rangle \,\langle 0|(\bar{\psi}
O_{1}\psi )|M_{2}\rangle\, .
\label{Factor-M-2}
\end{equation}

We want to investigate the large $N_{c}$ behavior of the pion distribution
functions. They are defined in terms of matrix
elements
\begin{equation}
\langle M_{1}|T\left\{ \left[ \bar{\psi}(z_{1}^{\prime })\Gamma
_{1}t^{B}\psi (z_{1})\right] \left[ \bar{\psi}(z_{2}^{\prime })\Gamma
_{2}t^{B}\psi (z_{2})\right] \right\} |M_{2}\rangle \, .
\end{equation}
We remind that
\begin{equation}
{\rm Sp}(t^{A}t^{B})=\frac{1}{2}\delta ^{AB} \, ,
\end{equation}
\begin{equation}
\sum\limits_{B}t_{ij}^{B}t_{kl}^{B}=\frac{1}{2}\left( \delta _{il}\delta
_{jk}-\frac{1}{N_{c}}\delta _{ij}\delta _{kl}\right) \, .
\end{equation}
In the large $N_{c}$ limit this reduces to
\begin{equation}
\sum\limits_{B}t_{ij}^{B}t_{kl}^{B}\rightarrow \frac{1}{2}\delta _{il}\delta
_{jk} \, .
\end{equation}
Hence for any color singlet matrices $\Gamma_{1,2}$ we have
\[
\langle M_{1}|T\left\{ \left[ \bar{\psi}(z_{1}^{\prime })\Gamma
_{1}t^{B}\psi (z_{1})\right] \left[ \bar{\psi}(z_{2}^{\prime })\Gamma
_{2}t^{B}\psi (z_{2})\right] \right\} |M_{2}\rangle
\]
\begin{equation}
\stackrel{N_{c}\rightarrow \infty }{\longrightarrow }-\frac{1}{2}\langle
M_{1}|T\left\{ {\rm Sp}\left[ \psi (z_{2})\otimes \bar{\psi}(z_{1}^{\prime
})\right] \Gamma _{1}\left[ \psi (z_{1})\otimes \bar{\psi}(z_{2}^{\prime })
\right] \Gamma _{2}\right\} |M_{2}\rangle\, .
\label{Factorization-M-2}
\end{equation}
We stress that here in $\bar{\psi}(z_{1}^{\prime })\otimes \psi (z_{2})$ the
tensor product refers to the spin flavor indices while the color indices are
contracted. The trace Sp also refers only to spin-flavor indices.

Applying the large $N_{c}$ factorization formula (\ref{Factor-M-2})
to the matrix element in the rhs of (\ref{Factorization-M-2}) one
obtains

\[
\langle M_{1}|T\left\{ \left[ \bar{\psi}(z_{1}^{\prime })\Gamma
_{1}t^{B}\psi (z_{1})\right] \left[ \bar{\psi}(z_{2}^{\prime })\Gamma
_{2}t^{B}\psi (z_{2})\right] \right\} |M_{2}\rangle _{connected}
\]
\[
\stackrel{N_{c}\rightarrow \infty }{\longrightarrow }-\frac{1}{2}{\rm Sp}
\left\{ \langle M_{1}|T\left[ \psi (z_{2})\otimes \bar{\psi}(z_{1}^{\prime })
\right] |0\rangle \,\Gamma _{1}\,\langle 0|T\left[ \psi (z_{1})\otimes \bar{
\psi}(z_{2}^{\prime })\right] |M_{2}\rangle \,\Gamma _{2}\right\}
\]
\begin{equation}
-\frac{1}{2}{\rm Sp}\left\{ \langle 0|T\left[ \psi (z_{2})\otimes \bar{\psi}
(z_{1}^{\prime })\right] |M_{2}\rangle \,\Gamma _{1}\,\langle M_{1}|T\left[
\psi (z_{1})\otimes \bar{\psi}(z_{2}^{\prime })\right] |0\rangle \,\Gamma
_{2}\right\}\, .
\label{factorization-2-3}
\end{equation}
Now we apply the general large $N_c$
eq.~(\ref{factorization-2-3}) to matrix
elements entering in definitions of twist-4 parton distributions
(\ref{T-T-tilde-def}). We obtain the following result for these
functions
\[
T_{qq}^{(+)}(x_{1},x,x_{2})\equiv \left[
T_{qq}(x_{1},x,x_{2})+\tilde{T}_{qq}(x_{1},x,x_{2})\right]
\]
\[
=-g^{2}\frac{F_{\pi }^{2}}{8}\left[ \delta (x_{1}-x_{2}-1)\phi _{\pi
}(x_{1}-x)\phi _{\pi }(x_{1})+\delta (x_{2}-x_{1}-1)\phi _{\pi
}(x_{2}-x)\phi _{\pi }(x_{2})\right]
\]
\begin{equation}
\times \left( -\frac{4}{9}\delta ^{ab}+\delta ^{a3}\delta^{b3}\right)\, .
\label{T-plus-plus-res}
\end{equation}
Here $\phi _{\pi }(x)$ is the pion
{\em twist-2}  distribution amplitude defined as
\begin{equation}
\frac{1}{F_\pi} \int \frac{d\lambda}{2\pi} e^{i x \lambda (P\cdot n)}
\langle 0| \bar \psi (\lambda n)  \hat n \gamma_5
\frac{\tau^b}{2}  \psi (0)
| \pi^a (P )  \rangle
=i \delta^{ab}\phi_\pi (x)\, .
\label{phi_pion}
\end{equation}
Here $F_\pi\approx 93$~MeV is the pion decay constant and $a,b$ are isospin indices.
Also it is easy to
see from definition eqs.~(\ref{T-plus-def},\ref{T-T-tilde-def})
that at leading order of the $1/N_c$ expansion
\begin{equation}
T_{qq}^{(-)}(x_{1},x,x_{2})=0\, .
\label{T-plus-minus-res}
\end{equation}

We see that in the large $N_c$ limit the meson twist-4 four-quark
distributions are expressible in terms of twist-2 meson
distribution amplitudes. Now we are in a position to compute the
twist-4 part of $F_2(x_B)$. We insert expressions (\ref{T-plus-plus-res},\ref{T-plus-minus-res})
into eq.~(\ref{F2-tw-4-corrected-start-factorized})
\[
F_{2}^{twist-4}(x_{B})=-g^{2}\frac{x_{B}^{2}}{Q^{2}}\,\frac{F_{\pi }^{2}}{4}
\left( -\frac{4}{9}\delta ^{ab}+\delta ^{a3}\delta ^{b3}\right)
\]
\[
\times \left\{ 2\phi _{\pi }(x_{B})\int dx_{1}\phi _{\pi }(x_{1})\frac{1}{
x_{1}}\right.
\]
\begin{equation}
\left. -\int dx_{1}dy_{1}\phi _{\pi }(y_{1})\phi _{\pi }(x_{1})\delta
(x_{1}-y_{1}-x_{B})\left[ \frac{1}{(1-y_{1})y_{1}}+\frac{1}{(1-x_{1})x_{1}}
\right] \right\}\, .
\end{equation}
From this equation we can easily compute pion isospin-2 structure
function
\begin{equation}
 F_{2}^{I=2}=\left[ F_{2}^{twist-4}\right] ^{\pi
^{+}}+\left[ F_{2}^{twist-4}\right] ^{\pi ^{-}}-2\left[ F_{2}^{twist-4}
\right] ^{\pi ^{0}}\, .
 \label{F-I2-def}
\end{equation}
Here we have taken into account that the twist-2 contributions are
cancelled in the rhs of eq.~(\ref{pervoe}).
The isospin factor of eq.~(\ref{T-plus-plus-res}) yelds
\[
\left( -\frac{4}{9}\delta ^{ab}+\delta ^{a3}\delta ^{b3}\right) ^{\pi
^{+}}+\left( -\frac{4}{9}\delta ^{ab}+\delta ^{a3}\delta ^{b3}\right) ^{\pi
^{-}}-2\left( -\frac{4}{9}\delta ^{ab}+\delta ^{a3}\delta ^{b3}\right) ^{\pi
^{0}}
\]
\begin{equation}
=\left( -\frac{4}{9}\right) +\left( -\frac{4}{9}\right) -2\left( -\frac{4}{9}
+1\right) =-2\, .
\label{tensor-I2}
\end{equation}
Therefore the final expression for $F_{2}^{I=2}(x_B)$
has the form
\[
F_{2}^{I=2}(x_{B})=g^{2}\frac{x_{B}^{2}}{Q^{2}}\,
\frac{F_{\pi }^{2}}{2}\left\{ \phi _{\pi }(x_{B})\int_0^1 dz
\frac{\phi _{\pi
}(z)}{z(1-z)}\right.
\]
\begin{equation}
\left. -\int_0^{1-x_B} dz\phi _{\pi }(z)\ \phi _{\pi }(x_{B}+z)
\left[ \frac{1}{z(1-z)}+\frac{1}{(x_{B}+z)(1-x_B-z)}
\right] \right\}\, .
\label{F-2-tw-4-general}
\end{equation}

\noindent
{\bf 4.}~Using the large $N_c$ relation (\ref{F-2-tw-4-general})
one can see that our result obeys the normalization
constraint
\begin{equation}
\int\limits_{0}^{1}dx_{B}\ F_2^{I=2}(x_B)=g^{2}
\frac{F_{\pi }^{2}}{2Q^{2}}
\label{M1}
\end{equation}
independently of the shape of pion distribution amplitude
 $\phi _{\pi }(x)$. We see that the first Mellin moment of
$F_2^{I=2}(x_B)$ in the large $N_c$
limit is given solely in terms of the pion decay constant $F_{\pi
}$ and not sensitive to the details of internal structure of the
pion. Recently this moment was computed in the lattice QCD \cite{Lattice}
with the results
\begin{eqnarray}
\label{latt}
\int\limits_{0}^{1}dx_{B}\ F_{2}^{I=2}(x_{B})&=&
0.27(10)\ g^{2}
\frac{F_{\pi }^{2}}{Q^{2}} \, .
\end{eqnarray}
This calculation gives about two times smaller result
for the first Mellin moment of $F_2^{I=2}(x_B)$
than our model independent large $N_c$ result.
Taking into account various approximations standing behind our
result (\ref{M1}) [leading order of the $1/N_c$ expansion] and
the lattice result (\ref{latt}) [quenched approximation, extrapolation
in light quark masses etc.] we find that the agreement between the two values
is quite reasonable.
Note that in contrast to the euclidean lattice calculations which can access
the Mellin moments of parton distribution our approach allows
us to compute the shape of the twist-4 distributions in
mesons.

Let us now compute the third Mellin moment of $F_{2}^{I=2}(x_{B})$.
 This moment in the large $N_c$ limit is sensitive
to the form of the pion distribution amplitude
$\phi_\pi(x)$. To demonstrate this we expand the pion distribution
amplitude in Gegenbauer series

\begin{eqnarray}
\phi_\pi(x)=6x(1-x)\biggl[1 +\sum_{n=2}^\infty a_n C_n^{3/2}(2 x-1)
\biggr]\, ,
\label{geg}
\end{eqnarray}
where $C_n^{3/2}(2 x-1)$ are Gegenbauer polynomials. It is a
matter of a simple calculation to see that the third Mellin moment
of $F_{2}^{I=2}(x_{B})$ is sensitive only to
the coefficient $a_2$ in the Gegenbauer expansion of the pion
distribution amplitude (\ref{geg})

\begin{equation}
\int\limits_{0}^{1}dx_{B}\ x_B^2
F_2^{I=2}(x_B)=g^{2}
\frac{F_{\pi }^{2}}{Q^{2}}\ \frac{3}{10}\biggl(1+a_2\biggr)\, .
\label{M3}
\end{equation}
Generically the $N$th Mellin moment (for odd $N$) can be
expressed in terms of Gegenbauer coefficients up to order $N-1$.

\begin{figure}[t]
\epsfxsize=12cm
\centerline{\epsffile{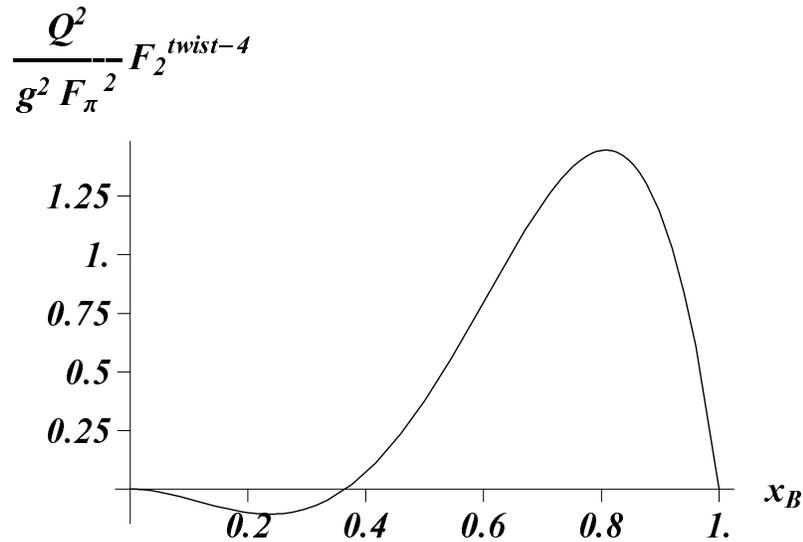}}
\caption{{\em The result for $\frac{Q^2}{g^2 F_\pi^2}
F_2^{I=2}(x_B)$ in the large $N_c$ limit for the case
of the asymptotic pion distribution amplitude.}}
\label{iaa}
\end{figure}

Using the general large $N_c$ expression (\ref{F-2-tw-4-general}) we
compute the twist-4 structure function $F_2^{I=2}(x_B)$
 for the asymptotic pion
distribution amplitude
\begin{equation}
\phi _{\pi }(x)=6x(1-x)\, .
\end{equation}
The result (see Fig.~\ref{iaa}) is given by

\begin{equation}
F_2^{I=2}(x_B)=6\frac{F_{\pi }^{2}}{Q^{2}}
\,g^{2}x_{B}^{2}(1-x_{B})(-1+2x_{B}+2x_{B}^{2})\, .
\end{equation}

We see that function
$F_2^{I=2}(x_B)$ is concentrated at
relatively large values of $x_B$. Let us note that the isospin-2
pion structure function should have at least one zero due to the general sum
rule

\begin{equation}
\int_0^1 \frac{dx_B}{x_B^2}\ F_{2}^{I=2}(x_{B})=0\, ,
\end{equation}
which directly follows from
eq.~(\ref{F2-tw-4-corrected-start-factorized}) and holds already at finite $N_c$.

We have demonstrated that in the large $N_c$ limit the higher
twist quark-antiquark correlations in  mesons are expressible
completely in terms of their twist-2 distribution amplitudes. This
model-independent result provides an illustration of general
relations between higher twist structure functions and hadron wave
functions as recently discussed by S.J.~Brodsky~\cite{Brodsky}.

\noindent
{\bf Acknowledgements}\\
We are grateful to A.~Belitsky, J.~Collins, M.~Franz,
L.N.~Lipatov, H.-Ch.~Kim, V.Yu.~Petrov, O.V.~Teryaev, and C.~Weiss for useful discussions.
The work is supported in parts by DFG, BMBF, and RFBR.

\end{document}